\newcommand{\format}[2]{#1}
\format{
  \documentclass[11pt,twoside,final]{article} 
  \usepackage{palatino}
}
{
  \documentclass[12pt,oneside]{article}
  \usepackage{setspace}
  \doublespacing
  \usepackage{palatino}
}
\usepackage{amssymb,amsmath}
\usepackage[round,colon,authoryear]{natbib} 

\newif\ifpdf \ifx\pdfoutput\undefined
\pdffalse 
\else
\pdfoutput=1 
\pdftrue \fi

\ifpdf \usepackage[pdftex]{graphicx} \else \usepackage{graphicx} \fi
\ifpdf \usepackage[pdftex]{hyperref} \else \usepackage[hypertex]{hyperref} \fi

\newcommand{\balpha}{{\boldsymbol{\alpha}}}
\newcommand{\bbalpha}{\bar{\boldsymbol{\alpha}}}

\newcommand{\bbF}{{\mathbb F}}

\newcommand{\bbR}{{\mathbb R}}

\newcommand{\bde}{\begin{description}}

\newcommand{\ben}{\begin{enumerate}}
\newcommand{\beq}{\begin{eqnarray}}

\newcommand{\beqn}{\begin{eqnarray*}}

\newcommand{\bJ}{{{\mathbf J}}}

\newcommand{\bqu}{\begin{quote}}

\newcommand{\bvb}{\begin{verbatim}}

\newcommand{\cA}{{\mathcal A}}
\newcommand{\cD}{{\mathcal D}}
\newcommand{\cF}{{\mathcal F}}

\newcommand{\cL}{{\mathcal L}}
\newcommand{\cN}{{\mathcal N}}
\newcommand{\cP}{{\mathcal P}}

\newcommand{\cZ}{{\mathcal Z}}

\newcommand{\ea}{\end{eqnarray}}

\newcommand{\ede}{\end{description}}
\newcommand{\een}{\end{enumerate}}
\newcommand{\eeq}{\end{eqnarray}}
\newcommand{\eeqn}{\end{eqnarray*}}

\newcommand{\equ}{\end{quote}}

\newcommand{\evb}{\end{verbatim}}

\newcommand{\<}{\langle}

\newcommand{\ZJ}{Z({\bf J})}
\newcommand{\ZJr}{Z_r({\bf J})}

\newcommand{\pmbeg}{\begin{pmatrix}}
\newcommand{\pmend}{\end{pmatrix}}

\renewcommand{\>}{\rangle}

\format{
\textheight 8.5 in 
\topmargin 0 in 
\textwidth 6 in 
\oddsidemargin 0in
\evensidemargin 0.5in 
\marginparwidth 1.3in
}{
\oddsidemargin=.375in
\evensidemargin=.375in
\textwidth=5.75in
\textheight=8.25in
\topmargin=0in}

\format{}{}
\format{}{  
  \renewenvironment{abstract}{%
    \begin{center}%
      {\bfseries \abstractname\vspace{-.5em}\vspace{-.5em}}%
    \end{center}%
    \quotation
  }{\endquotation}
}

\title{\vskip -20mm Fluctuation-dissipation theorem and models of learning}
\author{Ilya Nemenman\\ 
  Kavli Institute for Theoretical Physics\\ University of California,
  Santa Barbara, CA 93106\\
  and Joint Centers for Systems Biology, Columbia University\\
  1150 St Nicholas Ave, Rm 121, New York, NY 10032\\
  {\tt ilya.nemenman@columbia.edu}
}
\begin{document}
\maketitle

\ifpdf\DeclareGraphicsExtensions{.pdf, .jpg, .tif} \else
\DeclareGraphicsExtensions{.eps, .jpg} \fi

\renewenvironment{description} {\list{}{ \itemsep 1mm \parsep 0mm
    \itemindent -10mm \leftmargin 10mm }} {\endlist}
\newcommand{\bgm}[1]{\mbox{\boldmath $#1$}}
\newcommand{\bgms}[1]{\mbox{{\scriptsize \boldmath $#1$}}}
\newcommand{\ul}[1]{\underline{#1}} \newcommand{\bgt}[1]{{\boldmath
    $#1$}}

\begin{abstract}
  Advances in statistical learning theory have resulted in a multitude
  of different designs of learning machines. But which ones are
  implemented by brains and other biological information processors?
  We analyze how various abstract Bayesian learners perform on
  different data and argue that it is difficult to determine which
  learning--theoretic computation is performed by a particular
  organism using just its performance in learning a stationary target
  (learning curve).  Basing on the fluctuation--dissipation relation
  in statistical physics, we then discuss a different experimental
  setup that might be able to solve the problem.
\end{abstract}

\section{Introduction}

Learning based on experience (variously known as sensing, information
processing, or adaptation) is ubiquitous on all scales in biology.
For example, on the molecular scale, the Lac operon in {\em E.\ coli}
learns the lactose concentration to produce $\beta$--galactosidase
(the lactose--metabolizing enzyme) in proper quantities \citep{lac}.
Similarly, in the sensory system, the retinal phototransduction
cascade uses the information in the arrivals of photons to learn the
instantaneous light intensity and thus the current visual scene
\citep{detwiler-etal-00}. Additionally, it also learns the ambient
light level (to adapt to it) and the temporal correlations (to
estimate motion) \citep{reichardt,rob}. On the scale of cellular
(neuronal) networks, learning, memory, and adaptation in the neural
code are a text book knowledge [see e.~g.,
\citep{BrennerEtAl2000,fairhall-02}].  At yet larger scales,
experiments on rodents are revealing how they learn and respond to
changes in their environments \citep{gallistel-etal-01}; this is a
simple, albeit quantifiable, example of the general phenomenon we call
``learning'' in everyday life. Finally, we may also view evolution as
an example of learning, where entire species adapt to the world by
means of natural selection.

The creativity of theorists matches that of the Nature, and the number
of various learning paradigms different in their goals, assumptions,
methods, and performance guarantees is astonishing---too large to
enumerate here.  Fortunately, it is possible to build uniform
foundations for many of these {\em learning machines}
\citep{vapnik-98,nemenman-00,bnt-01}, and to find analogs among, say,
Structural Risk Minimization \citep{vapnik-98} and Bayesian
\citep{bayes} models.  However, while one might argue that biological
systems are (efficiently) implementing one of many abstract
learning--theoretic computations
\citep{attneave-54,barlow-59,barlow-61,atick-92,bnt-01}, it is often
unclear which exact computation is performed in a particular case.
For example, what is a learning--theoretic model equivalent to a rat
\citep{gallistel-etal-01}? or to a simple neural network that tries to
maximize its reward \citep{seung-03}? Answering such questions may
explain some of animal behaviors, uncover which assumptions they make
about the surrounding world, and establish quantitative limits on
their learning performance.

To attack the problem, one can construct a biologically plausible
computing machine with a known learning--theoretic equivalent
\citep{rao-04} and then search for a structural similarity with a real
leaving organism.  We do not pursue this approach, but choose a more
traditional route to establish the equivalence: comparison of
performance of real creatures to that of abstract learning machines.
As we will argue, analysis of paradigmatic {\em learning curves} is
not always easy. Thus one of the most important results of the paper
is a suggestion of a new protocol for making such comparisons.  The
intuition behind the suggestion comes from the famous
Fluctuation--Dissipation Theorem \citep{ma-85}. Based on this analysis
and on plausible assumptions about statistics of natural stimuli, we
also suggest that a particular learning--theoretic model might be
better suited for biological learning than the alternatives, and thus
it should be realized often in reality if optimization of learning is
desired.

To follow this route, we need to understand characteristics of
learning within different mathematical models fairly well, and a large
part of the paper is devoted to this. Analysis is done in the
framework of unsupervised Bayesian learning of probability
distributions since (a) evidently, Bayesian paradigm is relevant in
neuroscience \citep{kording-wolpert-04}, (b) as mentioned, different
learning frameworks are often equivalent, and (c) according to
\cite{bnt-01}, other learning problems usually can be reduced to
unsupervised learning of distributions. Much of this first part of the
paper is an abridged review, which follows the spirit and the notation
of \citep{bnt-01} and often prefers clarity to mathematical rigor.  We
do not try to make the review self-contained, but instead want to
elucidate and emphasize some important points that might have been of
a lesser interest in other contexts and also to present some novel
results, mostly developed in the Appendices.  After these
developments, we return to the main question of this work: how can one
realistically determine an equivalent learning--theoretic model for a
biological organism?

\section{The basics of learning}
\label{sec:basics}

Learning machines should be powerful enough to explain complex
phenomena. However, when data is scarce, this power leads to
overfitting and poor generalization. Thus a balance must be struck
between the abilities to explain and to overfit, and this balance will
depend on the amount of data available. In accord, much of statistical
learning theory
\citep{jeffreys-36,schwartz,janes-79,rissanen-89,clarke-barron-90,mackay-92,vijay-97,vapnik-98,nemenman-00,bnt-01}
has been devoted to putting the famous paradigm of William of Ockham,
{\em Pluralitas non est ponenda sine neccesitate}, on firm
mathematical footing in various theoretical frameworks. In particular,
in Bayesian formulation \citep{bayes,bayes2}, we know now how proper
Bayesian averaging creates {\em Occam factors} that punish for
complexity and weigh posterior probabilities towards those estimates
among a {\em finite} set of {\em parametric} model families that have
the best overall predictive power \citep{bnt-01}, but do not
necessarily produce the best fit to the observed data. This has been
called {\em Bayesian model selection}.\footnote{With the
  creationism--evolution tension mounting in teaching of biology in
  the U.\ S.\ schools, it is amusing to see how two friars, William of
  Ockham and Thomas Bayes, teamed up with modern day mathematicians to
  produce, in my view, the clearest formulation of the theory of
  learning from past experiences. If this approach results in a better
  understanding of biological designs, the situation will be even more
  peculiar.}

The waters get murkier in a {\em nonparametric} or {\em infinite
  parameter} setting when the whole functional form of an unknown
object is to be inferred.  Bayesian nonparametric developments
generally parallel parametric ones, and techniques of Quantum Field
Theory (QFT) help in computations
\citep{BCS,holy-97,bnt-01,cucker-smale-01,nb-02,lemm}.  However, the
exact relationship between the two settings is unknown, and some
results suggest subtle logarithmic differences between the cases
\citep{hall-hannan-88,rissanen-etal-92,bnt-01}.

We now review these and other Bayesian learning machines, and we start
with an introduction of some important and useful quantities.

Suppose we observe i.\ i.\ d.\ samples $x_i$, $i=1\dots N$. For
simplicity, we assume that $x$ is a scalar, but this does not affect
most of the discussion. We need to estimate the probability density
that generates the samples. A priori we know that this density,
$Q(x|\balpha)$, can be indexed by some (possibly infinite dimensional)
vector of parameters $\balpha$, and the probability of each parameter
value is $\cP(\balpha)$.  Then we define the {\em density of models}
({\em solutions}), at a given {\em distance} ({\em dissimilarity}, or
{\em divergence}) $D(\bbalpha,\balpha)=\epsilon$ away from the unknown
true target $\bbalpha$, which is being learned:
\begin{equation}
  \rho(\epsilon;\bbalpha) =\int d\balpha\, \cP(\balpha)\,
  \delta\left[D(\bbalpha,\balpha) - \epsilon\right]\,.
\end{equation}
For Bayesian inference of probability densities, the correct measure
of dissimilarity is the Kullback--Leibler divergence, $D_{\rm
  KL}(\bbalpha||\balpha) = \int dx\, Q(x|\bbalpha)\log\format{
  [Q(x|\bbalpha)/Q(x|\balpha)]}{\frac{Q(x|\bbalpha)}{Q(x|\balpha)}}$
\citep{bnt-01}, which has an important information--theoretic
interpretation \citep{ct-91}. However, in other situations different
choices of $D$ can and should be made.

Performance of a Bayesian learner is usually measured by the speed
with which the posterior probability concentrates for $N\to\infty$
(the {\em learning curve}) and by whether the point of concentration
is the true unknown target ({\em consistency}).  These characteristics
illuminate the importance of $\rho$, as it relates to both of them.
First, it has been proven that if, for $\epsilon\to+0$, the density,
$\rho(\epsilon;\bbalpha)$, is not zero, then the Bayesian problem is
consistent \citep{nemenman-00,bnt-01}.  Intuitively, this is because,
for large density, statistical fluctuations of the sample and of the
estimated parameters result in small $D_{\rm KL}(\bbalpha||{\rm
  estimate})$, making convergence to the target almost certain.

Relation of $\rho$ to the learning curve is more complicated.  We can
calculate the average (over samples) Occam factor for a given target
(the {\em generalization error}, or the {\em fluctuation determinant})
to the leading order in $1/N$:
\begin{equation}
  \cD(\bbalpha;N)\approx -
  \log \int d\epsilon\, \rho(\epsilon;\bbalpha) {\rm
    e}^{-N\epsilon}\,.
  \label{eq:D}
\end{equation}
This is the term that emerges as the penalty for complexity in
Bayesian model selection \citep{vijay-97,bnt-01}. If averaged over
$\bbalpha$, the Occam factor becomes {\em predictive information}
\citep{bnt-01}, which is the average number of bits that $N$ samples
provide about the unknown parameters,
\begin{equation}
  I_{\rm pred}(N) = \int d\bbalpha\, \cP(\bbalpha) \cD(\bbalpha;N)\,.
\end{equation}
Finally, one can define the {\em universal learning curve}, which
measures the expected $D_{\rm KL}$ between the target and the estimate
after $N$ observations \citep{bnt-01}. Up to the first order in the
large parameter $N$, this is\pagebreak[0]
\begin{align}
  \Lambda(\bbalpha;N) &\approx \frac{d \cD(\bbalpha; N)}{dN}\,,\label{eq:Lama}\displaybreak[0]\\
  \Lambda(N)&= \int d\bbalpha\,\cP(\bbalpha) \Lambda(\bbalpha;N) \approx
  \frac{d I_{\rm pred}}{dN}\,.\label{eq:Lam}
\end{align}

Many of these quantities, especially $I_{\rm pred}$, are also natural
objects when analyzing complexity of a time series \citep{bnt-01}.

\section{Different models of learning}
Since one of the goals of this work is to investigate if learning
machines can be discriminated by means of their learning performance,
specifically $\Lambda(N)$, here we discuss how $\Lambda$ depends on
$N$ for different scenarios.
 
\subsection{Learning in a finite set of parameters}
Consider a setup where $\balpha$ can take $M$ discrete values
$a_1,a_2,\format{}{$ $}\dots,a_M$ with a priori probabilities
$\cP_1,\cP_2,\dots,\cP_M$, and their divergences from the target $a_1$
are $0=d_1<d_2<\cdots<d_M$. The density is $\rho(\epsilon;a_1) =
\sum_{i=1}^M \cP_i \delta(d_i-\epsilon)$. For $N\to\infty$, we have
\begin{eqnarray}
  \cD(a_1;N) &=& - \log \sum_{i=1}^M \cP_i \exp [-Nd_i]
\approx -\log
\cP_1 -\cP_2/\cP_1 \exp[-Nd_2],\\ 
\Lambda(a_1;N) &\approx&
d_2\cP_2/\cP_1 \exp[-Nd_2].
\end{eqnarray}
So exponential learning curves (and asymptotically finite $\cD$ and
$I_{\rm pred}$) correspond to learning a possibility in a finite set.
Similarly, we can construct models with $\Lambda (N) \propto1/N^\nu,
\nu>1$, and they will also have asymptotically finite $I_{\rm pred}$.

\subsection{Finite parameter learning}
Now let the target probability density $Q(x|\bbalpha)$, or a {\em
  model}, belong to a set of densities $A$, a {\em model family}, that
can be indexed by a vector of parameters $\balpha\in A$, $\dim \balpha
= K<\infty$, and $\forall \balpha\in A,\, \cP(\balpha)>0$.  Then if
$A$ is not compact, or if the KL divergence between $\bbalpha$ and the
boundary of $A$ is larger than $\epsilon$, then the density of
solutions for such $K$--parametric family is \citep{bnt-01}\footnote{A
  different scaling dimension $d_K$ may appear in these formulas
  instead of $K$, the number of parameters. For example, for a
  redundant parameterization, $d_K<K$. Opposite situations, $d_K>K$
  and even $d_K\to\infty$, are also possible \citep{bnt-01}.}
\begin{equation}
  \rho (\epsilon;\bar\balpha) \approx
  {\cal P} (\bar\balpha|r) \frac{2 \pi^{K/2}}{\Gamma(K/2)} 
  \frac{ \epsilon^{(K-2)/2}}{\sqrt{\det \cF_{K}}}\,,
  \label{eq:rhoKpar}
\end{equation}
where
\begin{equation}
  \cF^{\mu\nu}_K(\bbalpha) = 
  \left.\frac{\partial^2 D_{\rm KL}(\bar\balpha||\balpha)}
    {\partial\alpha_\mu\partial\alpha_\nu}\right|_{{\balpha}
    ={\bar\balpha}}\,.
\end{equation}
$N\cF_K$ is the Fisher information matrix \citep{ct-91}; its
eigenvectors are the principal axes of the error ellipsoid in the
parameter space, and the (inverse) eigenvalues are variances of
parameter estimates along each of these directions. The prefactor $2
\pi^{K/2} / \Gamma(K/2)$ is the area of the $K$--sphere, and it has to
be multiplied by the fraction of the sphere that is inside $A$ if the
latter is (semi)compact.  Eq.~(\ref{eq:rhoKpar}) now gives
\begin{eqnarray}
I_{\rm pred}(N) &\approx&\cD(\bbalpha,N) \approx K/2 \log N,\\
\Lambda(N) &\approx& \Lambda(\bbalpha,N) \approx K/(2N).\label{lambdafin}
\end{eqnarray}

The situation changes slightly if $\bar Q \not \in A$ (here $\bar Q$
is the target density), and the prior assumptions about the world are
wrong. Then we find the best approximation to the target within $A$,
$\hat\balpha = \arg \min_{\alpha\in A} D_{\rm KL}(\bar Q||\balpha)$,
and define the distance between $\bar Q$ and $A$, $D_A(\bar Q) \equiv
D_{\rm KL}(\bar Q||\hat\balpha)$ [this is similar to the I-projection
\citep{csiszar-75}, but the order of arguments in $D_{\rm KL}$ is
different].  In this case, the model density is zero for $\epsilon\le
D_A(\bar Q)$, and the estimate concentrates near $\hat\balpha$ as
$N\to\infty$. Thus, if the radius of curvature of $A$ is much larger
than $\epsilon$, and $D_A(\bar Q)$ is also small, then
Eqs.~(\ref{eq:rhoKpar}, \ref{lambdafin}) generalize to
\begin{eqnarray}
  \rho(\epsilon;\bar\balpha) 
  &\approx&\left\{\begin{array}{l}
  {\cal P} (\hat\balpha) \frac{2 \pi^{K/2}}{\Gamma[K/2]} 
  \frac{ [\epsilon-D_A(\bar Q)]^{(K-2)/2}}{\sqrt{\det \cF_{K}}},\;
  \epsilon>D_A(\bar Q),\label{rhofinite}\\
  0,\; \epsilon\le D_A,
  \end{array}\right.\\
  \Lambda(\bar Q,N) &\approx& D_A(\bar Q) + K/(2N).\label{finlcshift}
\end{eqnarray}

\subsection{Nested finite parameter models}
\label{sec:finite}

Suppose now the target $\bar Q(x)$ that generates the observations
belongs to one of $R$ model families, $A_r, r=1\dots R$, with ${\rm
  Prob } ( \bar Q \in A_r) =\cP(r)$.  Models in each of the families
are indexed by parameters $\balpha^{(r)}$, $\dim \balpha^{(r)} =
K(r)<\infty$, so that the density of observing $x$ in a given model is
$Q_r(x|\balpha^{(r)})$.  Within each family, the parameters are a
priori distributed according to $\cP (\balpha^{(r)}|r)$.

We will assume that the families are {\em nested}. By this we mean
that $Q_r(x|\balpha^{(r)}) \format{}{$ $}\equiv Q(x|\balpha)$ are
independent of $r$, but that in each family the values of
$\alpha_\mu$, $\mu>K(r)$, are identically zero. Further, the nonzero
parameters have the same a priori distributions in all families:
\begin{eqnarray}
  \cP(\alpha_\mu|r) &=&\left\{ 
    \begin{array}{cc}
      p(\alpha_\mu)\,, & \mu\le K(r)\\
      \delta(\alpha_\mu)\,, & \mu>K(r)
    \end{array}
  \right.
  \label{eq:turn_on}
  \\
  \cP(\balpha|r) &=& \prod_{\mu=1}^R \cP(\alpha_\mu|r)\label{eq:pr_joint}
\end{eqnarray}
Thus a parameter $\alpha_\mu$ is ``switched on'' (or ``activated'')
when $r$ reaches $r_\mu \equiv \min_r \{r: K(r)\ge \mu\}$.  Discussion
of such nested models has been current in Bayesian
\citep{bayes2,bayes3} and frequentist \citep{regression} literature
for many years. However, we are unaware of any comprehensive analysis
relevant to important questions analyzed in our current presentation,
such as those in Appendices \ref{mfsel}--\ref{crossNestQFT}.

If $R\to\infty$, then we require that the union of all families forms
a {\em complete} set, so that every sufficiently smooth probability
density can be approximated arbitrarily closely by some member of the
union (if needed, this definition can be made more precise).

For simplicity, in this paper we focus on\footnote{Nestedness,
  completeness, and normality of the priors are needed only for
  comparison with models discussed later, and they are not essential
  for Bayesian learning.}  \pagebreak[0]
\begin{align}
  p(\alpha_\mu) &= \cN(0,\sigma_\mu^2)\,,
  \label{eq:pr_gauss}
  \displaybreak[0]
  \\
  \sigma_\mu&=c r_\mu^{-\beta},\, \beta\ge 0,\, c={\rm
    const}\,,
  \label{eq:pr_var}
\end{align}
where $\cN(a,b)$ denotes a normal distribution with the mean of $a$
and the variance of $b$.  In particular, $\beta=0$ corresponds to the
same in--family a priori variances for all active parameters.  This is
common when discussing Bayesian model selection.

While these priors describe a set of parametric models, another view
is also possible.  The joint distribution of $r$, $\balpha$, and
$\{x\}$ is $P(\{x\},\balpha,r) = Q(\{x\}|\balpha)\format{}{$
  $}\cP(\balpha|r)\cP(r)$, which results in
\begin{eqnarray}
  P(\{x\},\balpha) &=& \sum_r Q(\{x\}|\balpha)\cP(\balpha|r)\cP(r) 
  = Q(\{x\}|\balpha)\sum_r \cP(\balpha|r)\cP(r)\nonumber\\
  &\equiv& Q(\{x\}|\balpha)\cP(\balpha)\,,\label{eq:cPjointne}
\end{eqnarray}
where the last equation defines $\cP(\balpha)$, the overall prior over
$\balpha$. Unlike $\cP(\balpha|r)$, $\cP(\balpha)$ is not factorizable
and is not differentiable at zero for any $\alpha_\mu$, $\mu>K(1)$.
Thus the nested setup may be viewed as inference in a combined model
family with $K(R)$ parameters.  In particular, for $R$ and
$K(R)\to\infty$, the learning problem has a countable infinity of
parameters leading to the common assumption of equivalence with the
nonparametric inference.

It is of interest to calculate the combined a priori mean and variance
of $\balpha$. Integrating over all $\alpha_\nu$, $\nu\neq \mu$, we get
the combined prior for $\alpha_\mu$
\begin{equation}
  \cP(\alpha_\mu) = \delta(\alpha_\mu) \sum_{r<r_\mu}\cP(r) +
  p(\alpha_\mu)\sum_{r\ge r_\mu} \cP(r)\,.
\end{equation}
By Eq.~(\ref{eq:pr_gauss}), the a priori means of all parameters are
zero, and the variances are \pagebreak[0]
\begin{equation}
  \<\delta \alpha^2_\mu\> = \sigma^2_\mu \sum_{r\ge r_\mu} \cP(r).
  \label{eq:finite_var}
\end{equation}
Thus the bare variance $\sigma^2_\mu$ is ``renormalized'' by the
probability to be in a family, in which the parameter is nonzero.  An
interesting special case is
\begin{eqnarray}
  \cP(r) &\propto& r^{-\gamma}\,,\quad \gamma>1,\;\;R\to\infty,\\
  r_\mu &=& \mu\,.\label{eq:cPr}
\end{eqnarray}
Then the a priori variance gets a simple form
\begin{equation}
  \<\delta \alpha^2_\mu\> \propto \mu^{-\beta}
  \sum_{r=\mu}^{\infty}r^{-\gamma} \sim \mu^{-\beta-\gamma+1}\,.
  \label{eq:finite_varscale}
\end{equation}
Thus $\<\delta\alpha_\mu^2\>$ depends as much on the bare variance as
on the speed of decay of $\cP(r_\mu)$. This suggests that the learning
properties of the nested setup will depend equivalently on $\beta$ and
$\gamma$. In fact, as shown in Appendix \ref{mfsel}, this is not true:
while behavior of $p(\alpha_\mu)$ is important, any reasonable choice
of $\cP(r)$ does not effect success of the learning.

From Eq.~(\ref{rhofinite}) we can now evaluate the model density and
the learning curve for the nested setup.
For each value of $\bar\balpha$ and $r$, we can find the model
$\hat\balpha_r = \arg \min_{\alpha\in A_r} D_{\rm
  KL}(\bbalpha||\balpha)$ that best approximates $\bbalpha$ in
$A_r$, and define $D_r(\bbalpha) \equiv D_{\rm
  KL}(\bbalpha||\hat\balpha)$, the distance between $\bbalpha$ and
$A_r$. If $\bbalpha \in A_r$, then $\bbalpha=\hat\balpha$, and
$D_r(\bbalpha)=0$.  However, if $\bbalpha \not\in A_r$, then
$D_r(\bbalpha)>0$. We then have:
\pagebreak[0]
\begin{equation}
  \rho(\epsilon;\bar\balpha)
  =\sum_{r:\,D_r(\bbalpha)\le\epsilon} \cP(r)
  {\cal P} (\hat\balpha_r|r) \frac{2 \pi^{K(r)/2}}{\Gamma[K(r)/2]} 
  \frac{ [\epsilon-D_r(\bbalpha)]^{[K(r)-2]/2}}{\sqrt{\det \cF_{K(r)}}}\,.
  \label{eq:rhofinsum}
\end{equation}

The learning curve in this scenario strongly depends on the target.
Let $\bbalpha$ have $\bar r$ active modes (with $\bar r$ determined
according to $\cP(\bar r)$), and let each of these modes have an
amplitude $\sim\sigma$ (that is, $\beta=0$).  Then $D_r(\bbalpha)$ is
either exactly zero (for $r\ge \bar r$), or large (for $r<\bar r$).
So Eq.~(\ref{eq:rhofinsum}) becomes
\begin{equation}
  \rho(\epsilon\to0;\bbalpha_{{\rm typ}; \bar r}) \sim
  \sum_{r\ge\bar r} \cP(r)
  {\cal P} (\hat\balpha_r|r) \frac{2 \pi^{K(r)/2}}{\Gamma[K(r)/2]} 
  \frac{\epsilon^{[K(r)-1]/2}}{\sqrt{\det \cF_{K(r)}}},
\end{equation}
where $\bbalpha_{{\rm typ};\bar r}$ is a distribution typical in
$A_{\bar r}$.  This is dominated by $r=\bar r$, and for $N\gg K (\bar r)$
the learning curve is
\begin{equation}
  \Lambda(N) \approx
      \frac{K(\bar r)}{2N}.
\label{eq:Lnest}
\end{equation}
It is now clear that averaging over $\cP(\bar r)$ is not very
informative. Note also that, for $N\lesssim \bar r$, the learning
curve goes through a cascade of $K(r)/N$ behaviors, $1\le r\le\bar r$,
and changes of the prefactor of the $N^{-1}$ scaling correspond to
activations of new parameters, which happen rather abruptly
(cf.~Appendix \ref{mfsel}).

\subsection{Nonparametric learning} 
\label{sec:nonparam}

Nonparametric learning usually refers to inferring a functional form
of a probability density $Q(x)$, or rather of $\phi(x) \equiv -\log
Q(x)$, with some smoothness constraints on it. The constraints may be
in the form of bounding some derivatives of $Q$ or $\phi$, which was
the choice of \citet{hall-hannan-88} and
\citet{rissanen-etal-92}.\footnote{These authors used histogramming
  density estimators, which have no hierarchy of model families; this
  is especially true for \citet{rissanen-etal-92}, who allowed locally
  varying bin widths.  Therefore, these techniques can not be referred
  to as nested parametric methods. On the other hand, they allow an
  arbitrarily precise fit to any probability density and may require
  an arbitrarily large number of break points and density values for
  complete specification. This is the reason for treating them as
  nonparametric.}  Alternatively, in the Bayesian framework followed
here, the constraints may be incorporated into a functional prior that
makes sense as a continuous theory, independent of discretization of
$x$ on small scales. For $x$ in one dimension, the minimal and the
most common choice is \citep{BCS,aida-99,nb-02,lemm}
\begin{equation}
\label{eq:npprior}
{\cal P}[\phi(x)]= \frac{1}{\cZ}\exp \left[-\frac{\ell^{2\eta -1}}{2}\int
dx
\left(\frac{\partial^{\eta} \phi}{\partial x^{\eta}}
\right)^2\right]\, \delta \left[\frac{1}{l_0}\int dx\, {\rm
e}^{- \phi(x)} -1 \right],
\end{equation}
where $\eta>1/2$, $\cZ$ is the normalization constant, and the
$\delta$-function enforces normalization of $Q$.  The hyperparameters
$\ell$ and $\eta$ are called the smoothness scale and the smoothness
exponent, respectively. Fractional order derivatives are defined by
multiplying by the wave number to the appropriate power in the Fourier
representation of $\phi$ (we assume periodicity on $[0,1)$). 

This prior is equivalent to specifying a 1--dimensional Quantum Field
Theory \citep{BCS,holy-97,nb-02,lemm}, and QFT methods have been
successful in the analysis.  In particular, the maximum likelihood
estimate of the distribution, $Q^*(x)\equiv\exp[-\phi^*(x)]$, is given
by the following differential equation
\begin{equation}
  \ell^{2\eta-1} \bbR_{-2\eta}\frac{\partial^{2\eta}\phi^*(x)}{\partial
    x^{2\eta}}
  -N Q^*(x) +\sum_i \delta(x-x_i) = 0\,,\label{eq:npsp}
\end{equation}
where the operator $\bbR_{\theta}$ shifts the phase of each Fourier
component of its argument by $\pi\theta/2$ \footnote{For a
  comprehensive treatment of fractional differentiation the reader is
  referred to \cite{samko-87}. In particular, the action of the phase
  shift operator $\bbR_{\theta}$ may be calculated by the Wiener--Hopf
  method.}. The equation shows that derivatives of $\phi^*$ and $Q^*$
of order $2\eta-1$ have step discontinuities.  Thus, for $2\eta=1$ the
classical solution itself, $\phi^*(x)$, is discontinuous, and for
$2\eta<1$ the singularities are even more severe.  We may characterize
sample--dependent fluctuations in $Q^*$ by $D_{\rm KL} = \int dx \,
Q^*_1(x)\log Q^*_1(x)/Q^*_2(x)$, where $Q^*_1$ and $Q^*_2$ are saddle
point solutions for different sample realizations.  If $Q^*$ has, at
least, step discontinuities at the sample points, and these points are
random, then $D_{\rm KL}$ does not fall to zero as $N$ grows.
Therefore, the QFT setup becomes inconsistent at $\eta=1/2$, even
though Bayesian formulation is proper, and the prior can still be
normalized by, for example, going to the Fourier representation.
This is in contrast to the 
nested setup, where normalizable priors guarantee consistency.

\citet{BCS,bnt-01} have calculated the $\epsilon\to0$ model density
and the fluctuation determinant for different $\eta$'s.  By noticing
from Eq.~(\ref{eq:npsp}) that $N$ and $\ell$ can enter the solutions
only in a combination $N/\ell^{2\eta-1}$, we extend their results and
recover correct dependence not only on $\eta$ (for $\eta>1/2$), but
also on $\ell$:
\begin{eqnarray}
  \rho (\epsilon;\bar\phi) &\approx& A[\bar\phi] \,\epsilon^\xi \exp\left[
    -\frac{B[\bar\phi]}{\ell\epsilon^{1/(2\eta-1)}}\right],
  \label{eq:rhonp}\\
  \cD(\bar\phi;N) &\approx& C[\bar\phi]\,
  \left(\frac{N}{\ell^{2\eta-1}}\right)
  ^{1/2\eta}, \label{eq:Dnp} \\
  \label{qftlc}
  \Lambda(\bar\phi,N) &\approx& \frac{C[\bar\phi]}{2\eta\,\ell^{2\eta-1}}
  \left(\frac{N}{\ell^{2\eta-1}}\right)^{1/2\eta -1}.\label{lcqft}
\end{eqnarray}
Here $\xi$ depends only on $\eta$, and $A$, $B$, and $C$ are some
known related functionals that do not depend on $\ell$.  These
asymptotics kick in when $N\gg 1/\ell$. In particular, for smaller
$N$, $\Lambda \sim 1$ and is barely decreasing.  The dependence of
$C[\bar\phi]$ on $\bar\phi$ may be significant (and, possibly,
diverging for ill-behaved targets).  However, from
Eq.~(\ref{eq:rhonp}), the dependence on $\eta$ near $2\eta-1\to+0$ is
easier to analyze:
\begin{equation}
  C[\bar\phi] \sim (2\eta-1)^{1/2\eta -1}\,\label{eq:C1}
\end{equation}
with an undetermined value at $2\eta=1$. So for $\eta\to 1/2$, $\cD$
approaches extensivity in $N$, and then becomes ill-defined, again
signaling inconsistency. As discussed by \cite{bnt-01}, problems with
$\cD(N)/N \to {\rm const}$ are the most complicated correctly posed
learning problems that exist, and they can be studied in the Bayesian
QFT setting.

For comparison with the nested case (cf.~Appendix \ref{nmfourier}), we
may replace $\phi$ by its Fourier series,
\begin{eqnarray}
  \phi(x|\balpha)&\equiv&  - \log Q(x|\balpha) = \alpha_0
  +\sum_{\mu=1}^r
  \left(\alpha_\mu^+ \cos 2\pi\mu
    x + \alpha_\mu^- \sin 2\pi\mu x\right),\label{eq:fourier}\\
  \alpha_0 &=& \log \int dx \, \exp\left[-
    \sum_{\mu=1}^r \left(\alpha_\mu^+ \cos 2\pi\mu x + \alpha_\mu^-
      \sin 2\pi\mu x\right)\right]\,,\label{eq:a0}
\end{eqnarray}
with $r\to\infty$ (finite $r$ results in a finite parameter model).
The last equation enforces normalization, $\int dx\,Q(x|\balpha)=1$,
and it is equivalent to the constraint $\delta\big(\int
\exp[-\phi(x|\balpha)] dx -1\big)$ in the prior $\cP(\balpha|r)$ or
$\cP[\phi(x)]$. Since the Jacobian of the transformation $\phi(x)\to
\{\alpha^{\pm}_\mu\}$ is a constant, Eq.~(\ref{eq:npprior}) amounts to
zero-mean Gaussian priors over $\alpha_\mu^\pm$ with the variance
\citep{nb-02}
\begin{equation}
  \<(\delta \alpha_\mu^\pm)^2\>=\frac{2}{\ell^{2\eta-1}} 
  \frac{1}{\left(2\pi \mu\right)^{2\eta}},\quad\quad \mu>0.\label{eq:qft_var}
\end{equation}

Equa\-tions (\ref{eq:finite_varscale}, \ref{eq:qft_var}) suggest that
the nested and nonparameteric case are similar: the a priori means of
the amplitudes are zero, and the variances fall off as power laws in
$\mu$.  However, the field theory model requires the variance to
decrease at least as fast as $1/\mu$ (recall that $\eta>1/2$), while
the finite parameter case does not impose such constraints.  This is
an indication of an essential difference between the models: in the
nested case, the priors, specifically the a priori variances of
parameters, have less of an influence on learning. This can be easily
explained.  QFT nonparametric models do not have a sharp separation
between active and passive modes. The modes with low $\mu$ are
determined by the data, but fluctuations for larger $\mu$ are
inhibited only due to the small a priori variances,
Eq.~(\ref{eq:qft_var}). The exact attenuation of the fluctuation
depends on the values of $\eta$ and $\ell$, and the cumulative
contribution to posterior variance of the estimator may be
substantial. In contrast, for the finite parameter nested case, once
the most probable model family is determined, fluctuations of the
higher order parameters are inhibited {\em exponentially}
(cf.~Appendix~\ref{mfsel}).  The cumulative fluctuations are then
small and almost independent of the a priori parameter variances, and
the learning may succeed even for $\cP(r)$ with a long tail.

The dependence of the QFT model on the prior can be weakened by
treating $\ell$ as an unknown random variable and averaging over it
\citep{BCS,nb-02} (similar averaging over $\eta$ has not yet been
performed).  This is akin to nesting of finite
dimensional models 
and improves learning curves for a wide range of targets.  On the
other hand, integration over $\ell$ produces the theory that is not
necessarily local in $\phi(x)$, couples all of the Fourier amplitudes,
and is difficult to compare to the nested finite parameter setup
directly.  Therefore, we do not discuss the averaging in what follows,
but assume that the values of $\eta$ and $\ell$ used for learning are
the best for a particular target being learned.

\subsection{Comparing the performance}
\label{sec:comp}

One of the goals of the paper is to decide if learning curves can be
used to distinguish which learning machine is a good description of a
particular biological system. To this extent, we need to analyze
responses of various learners to data that they are not expecting.
Thus in this section we derive learning curves for a finite parameter,
a nested ($\beta=0$), and a QFT machine on data that is typical in the
prior of one of the two others. With mismatched data and expectations,
the learning curve can not be optimal, but may come quite close.

First, consider $\bar Q$ taken from the QFT prior. The learning curve
for the finite parameter model is given by Eq.~(\ref{finlcshift})---a
$N^{-1}$ decay towards some approximate target.  Further, as shown in
Appendix \ref{crossNestQFT}, the learning curves for complete nested
models, Eq.~(\ref{eq:Lcross}), and for the QFT machine,
Eq.~(\ref{lcqft}), which is the best possible machine for such data,
differ only logarithmically.

If instead we study a distribution that is typical in the nested case
for some $\bar r$ (equivalently, a finite parameter distribution with
$K(\bar r)$ parameters) then a finite parameter model again gives
Eq.~(\ref{finlcshift}). On the other hand, for $N\lesssim K(\bar r)$,
no complete learning machine can estimate all required unknown
parameters, and $\Lambda$ does not have a well defined scaling
\citep{nb-02}.  The differences between the machines emerge for $N\gg
\bar r$. The nested machine eventually asymptotes to
Eq.~(\ref{eq:Lnest}), and starts learning at the rate of $1/N$. However,
the QFT setup performs differently
: when $\Lambda\to0$ and all $\bar r$ modes are well approximated, the
machine continues trying to fit higher order modes, which it expects
to be present even though they are not. This will result in the same
fluctuation determinant as in Eq.~(\ref{eq:Dnp}), switching to the
usual asymptotic $\Lambda \propto (N\ell)^{1/2\eta -1}$ instead of

So, surprisingly, when the target has a finite number of degrees of
freedom, the nested setup is qualitatively faster than the QFT
learning machine!

\section{Learning a changing target}
\label{sec:change}

One never needs to know the distribution that generated the data to an
infinite precision, and some $\epsilon>0$ approximation is usually
enough.  Further, if learning in biological systems is {\em
  stochastic}, as argued, for example, by \cite{seung-03}, then
$\epsilon$ is bounded from below by the noise variance. As shown by
\cite{fairhall-02} and especially by \cite{gallistel-etal-01},
convergence to the ``good enough'' estimate happens so quickly, that
the transient learning curves are difficult to resolve. Is then the
performance difference between the nested and the QFT scenarios seen
in the previous section important? And can it be used to discriminate
between the models?

\subsection{Model density and variable stimuli}

Notice that often the target itself changes while being learned. The
ambient light intensity may be fluctuating while our eye estimates it,
or the variance of angular velocities measured by a fly motion
sensitive neuron can be varied by an experimenter while the fly tries
to adapt to it 
\citep{fairhall-02}. In these cases one has to learn constantly to
stay at the allowed $\epsilon$--error, and then a faster learning
machine may be truly advantageous. However, even for a variable
target, the nested learner will not be helpful if (a) a small change
of the target parameters throws it back to a very large $\Lambda$, or
(b) the changing target may drift to a region where $\bar r$ is so
large that the nested setup is not better than the nonparametric
anymore.

\begin{figure}
  \centerline{\includegraphics[width=4.5in]{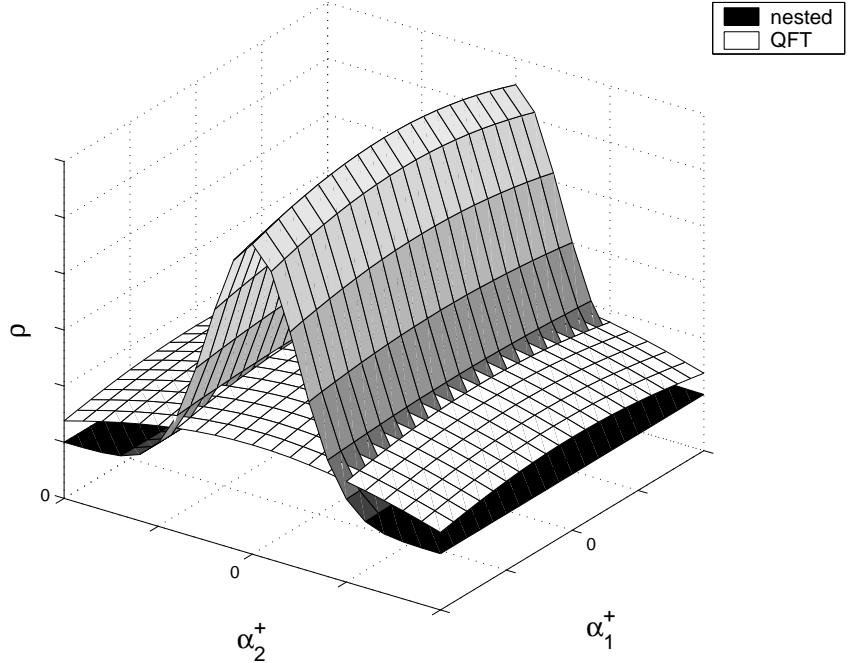}}
  \caption{\label{fig:density}Schematic density of models as a function of the
    target location.}
\end{figure}

To answer these concerns, instead of focusing on the density of
solutions as a function of the allowed error $\epsilon$, we will keep
$\epsilon$ fixed and vary $\bbalpha$.  For some small $\epsilon$, a
schematic drawing of dependence of $\rho$ on $\bar\alpha^+_1$ and
$\bar\alpha^+_2$ with the other parameters fixed at 0 is shown on
Fig.~\ref{fig:density}. In the nested case, there is a ridge along
$\alpha^+_2\approx0$, where the density is, at least,
$\sim1/\sqrt{\epsilon}$ larger than anywhere else,
cf.~Eq.~(\ref{eq:rhofinsum}).  The ridge comes from the prior,
Eq.~(\ref{eq:cPjointne}), for $\bar\alpha^+_2=0$ being singularly
larger than for $\bar\alpha^+_2\neq0$, and the singularity is then
smoothed out by $\epsilon$--approximation.  In comparison, the
nonparametric prior has a bivariate normal shape, which after
$\epsilon$--smearing results in a weak target dependency of
$\epsilon$--independent prefactors in Eq.~(\ref{eq:rhonp}); thus
$\rho(\bbalpha)$ varies slowly.  \footnote{The plots of
  $\cP(\bbalpha)$ and $\rho(\epsilon;\bbalpha)$ have very different
  meanings. The volume under the $\cP(\bbalpha)$ surface is fixed by
  normalization, $\int d\bbalpha \,\cP(\bbalpha)=1$. Thus high a priori
  probability on any singular line, e.~g., $\balpha_2^+=0$,
  necessarily means a lower prior elsewhere. Such considerations are
  the reason for {\em no free lunch} theorems \citep{wolpert-95}. In
  the language of the model density, the normalization condition is
  $\int d\epsilon\, \rho(\bbalpha;\epsilon) = 1$.  However, there are
  no constraints on the density integrated over $\bbalpha$, and a
  large density for some target does not necessarily result in a lower
  density elsewhere.}

Figure~\ref{fig:density} answers both of the concerns mentioned above.
For a QFT machine, densities everywhere are comparatively small. So a
small change of the target means vast and slow relearning.  In
contrast, if, for a nested case, $\bbalpha$ is in the large density
region, then there are many other models in the vicinity.  Small
parameter changes likely leave the target close, and not much needs to
be relearned.  Further, since the ridge drops off smoothly, models in
the vicinity of a large density target also have large densities, and
thus are learned fast as well. Of course, this holds only when the
target, indeed, varies mostly along a small set of directions, and
density ridges are aligned with those.  Importantly, since at a finite
$\epsilon$ the ridge has a finite width, a perfect alignment is {\em
  not necessary}.

We believe that many natural signals have such structure.  For
example, in phototransduction, instantaneous intensity is determined
by the statistics of reflectivities of objects that come in the view
and by the mean ambient light intensity. The statistics barely change
over long time scales, while the mean intensity depends on, for
example, clouds shading the sun and varies a lot and rapidly. The
photoreceptor may want to adapt to intricate details of the
distribution of reflectivities, but only after it accurately learns
the mean light level.  A similar separation of time scales is observed
in transcriptional regulation, where, for example, changes in the
lactose concentration happen on the scale of minutes, while statistics
of lactose bursts depends on the environment and is constant for
generations. In neuroscience, when estimating an angular velocity, a
fly takes into the account the preceding velocity variance
\citep{fairhall-02}, but it may not have time for reaction to higher
order moments.  Thus we believe that many natural learners that have a
need to learn fast, but also to be able to learn a very wide class of
models accurately on longer time scales, will be organized as nested
learning machines with the density ridges approximately adjusted to
fast variable directions.

\subsection{Fluctuation-dissipation and determining the model}

The prediction in the last Section brings us back to the main question
of this work: how can an underlying learning--theoretic computation be
inferred?  For many reasons, analysis of learning curves is not always
a good idea.  First, learning may happen so fast that resolving it
might present a problem \citep{gallistel-etal-01}. Second, to estimate
$\Lambda(N)$ reliably, we need to average, and a complete instance of
the learning curve is just one sample. Such averaging may require
prohibitively long experiments. Third, it is well know that animals
adapt. Thus eliciting the same response to the same target requires
large inter--trial time delays, further increasing the experimental
duration.  These problems can be traced to learning being an
inherently {\em transient} behavior, and they might become less severe
if we can characterize learning machines by some {\em stationary}
response properties. A hint comes from the Fluctuation--Dissipation
Theorem in statistical physics \citep{ma-85}, which states that, if a
system fluctuates in the presence of a linear dissipative restoring
force, then the variance of fluctuations (a stationary property) is
linearly related to the dissipation coefficient (a feature of the
transient response).  In our case, we may hope that response to a
variable target (fluctuations) reveals information about the learning
curve (dissipation).

In view of this suggestion, let us now analyze a few examples of a
variable target learning.\footnote{It is clear that stretching the
  theory of learning a fixed target to the fluctuating case may hide
  many potential pitfalls. We do this because we are unaware of any
  comprehensive treatments of the latter problem [though some progress
  is being made, cf.~\cite{deweese-98,atwal-04}]. }  We now denote by
$\balpha$ an estimate of $\bbalpha$ averaged over many presentations
of the same data. We keep almost all parameters fixed (or changing
very slowly), while $\bar\alpha_1(t)$, which is approximately the
direction of the ridge in the density of solutions, is allowed to
vary. If data are observed for a long time, then
$\bar\alpha_\mu\approx\alpha_\mu$ for $\mu\neq1$ (provided
$\bbalpha=\hat\balpha$). Now remember that $\Lambda$ is the expected
Kullback--Leibler divergence between $\bbalpha$ and $\balpha$, which
converges to the $\chi^2$ distance when it is small. Thus if
$\bar\alpha_1-\alpha_1$ is not large,
\begin{equation}
  \Lambda \propto (\alpha_1-\bar \alpha_1)^2.\label{eq:Lx}
\end{equation}

For a fixed target and $\Lambda\to0$ (that is, for $N\to\infty$,
$\hat\balpha = \bbalpha$), all learning curves we studied can be
summarized as
\begin{equation}
  \frac{\partial \Lambda}{\partial N} = -\zeta_N
  \Lambda^{\nu}.\label{eq:diffeqL}
\end{equation}
Here, in particular, $\nu=1$ corresponds to a finite set of solutions
along the direction of $\alpha_1$, $\nu=2$ is the finite--parameter or
nested case, and $\nu=3$ is the $\eta=1$ QFT model. In principle,
other values of $\nu\in(0;\infty)$ are possible. 
The constant $\zeta_N\sim1$ depends on the details of the learning
setup.  For example, for parametric cases, $\zeta_N=2/K(\bar r)$. 

For Eq.~(\ref{eq:diffeqL}), which is manifestly true for a fixed
$\bbalpha$, to also hold in the fluctuating target case, the learning
machine must quickly notice the target's variation and disregard old
samples as soon as they become outdated.  \cite{gallistel-etal-01}
show that a rat reacts to changes in the reward rates as fast an ideal
detector would.  Therefore, this assumption is reasonable for
biological systems.\footnote{We leave aside important comments by
  \citet{deweese-98}, who argued that time needed to notice a change
  may be not invariant with respect to the direction of the change.  }

If measurements are taken at a fixed rate, so that $dN/dt = {\rm
  const}$, we can combine Eqs.~(\ref{eq:Lx}, \ref{eq:diffeqL}) to get
\begin{equation}
  \frac{d \Delta}{dt} = -\zeta \, {\rm sign }(\Delta) \left|\Delta\right|^{2\nu-1} - v_{\bar \alpha}\,, \label{eq:dD}
\end{equation}
where $\Delta = \alpha_1 -\bar \alpha_1$ is the average error of the
estimation, $\zeta$ is some unknown constant with the dimensionality
of $1/t$ and is basically the scaled sampling rate, and $v_{\bar
  \alpha}$ is the drift velocity of the target.  Equation
(\ref{eq:dD}) is a clear example of a dissipative system, and it has
many analogues in the theories of classical and quantum dissipation
\citep{qdiss}. Note also that, unlike in the fluctuation--dissipation
analysis in statistical physics, the spectrum of fluctuations,
$v_{\bar \alpha}$, is not necessarily white and can be controlled by
an experimentalist, potentially providing more ways to probe the
underlying dissipative dynamics.

If the target's variation cannot be learned (incomplete or mismatched
machine), then Eq.~(\ref{eq:dD}) still holds.  However, because of
Eq.~(\ref{finlcshift}), we now have $\Delta=\alpha_1-\hat\alpha_1$
(recall that $\hat\balpha$ is the best approximation to the target by
a particular learning machine). Thus to trace the evolution of
$\alpha_1$ using Eq.~(\ref{eq:dD}), one would need to evaluate
$\hat\balpha(\bbalpha)$, which can be done from the stationary target
analysis.  Further, if the target varies along many learnable
directions, then for each such direction we have an analog of
Eq.~(\ref{eq:dD}), possibly with different $\xi$. So the dynamics of
$\Lambda$ is still given by Eq.~(\ref{eq:diffeqL}) with forcing, but
the dissipation constant depends on the number of varying parameters.

Let's now consider a few different examples of $v_{\bar \alpha}$.  If
$v_{\bar \alpha}=\cA$ is a constant, then asymptotically for
$t\to\infty$, setting $d \Delta/dt=0$, we find
\begin{equation}
  \Delta\to\Delta_\infty =- \left(\frac{\cA}{\zeta}\right)^{1/(2\nu-1)}\,.
  \label{eq:Dconst}
\end{equation}
The ratio $v_{\bar\alpha}/\zeta$ must be $\ll 1$, otherwise $\Delta$
is outside of the $\Lambda\to0$ asymptotic, for which
Eq.~(\ref{eq:diffeqL}) is valid. Thus, for small drifts, setups with
smaller $\nu$ win qualitatively.

It is also of interest to consider the situation when $\bar \alpha_1$
undergoes a Brownian motion, $\<v_{\bar
  \alpha}(t)v_{\bar\alpha}(t')\>=\Omega\, \delta(t-t')$. Writing the
Fokker-Planck equation for this Langevin dynamics, we easily find the
stationary distribution of $\Delta$,
\begin{equation}
  P(\Delta) =
  \frac{\nu}{\Gamma\left(\frac{1}{2\nu}\right)}\left(\frac{\zeta}{\nu\Omega}\right)^{1/(2\nu)}
  \exp \left\{- \frac{\zeta |\Delta|^{2\nu}}{\nu \Omega}\right\}, 
\end{equation}  
which results in the rms fluctuations of
\begin{equation}
  \Delta_{\rm rms} = \left\{\nu^{1/\nu}
    \frac{\Gamma\left(\frac{3}{2\nu}\right)}
    {\Gamma\left(\frac{1}{2\nu}\right)}\right\}^{1/2}
    \left(\frac{\Omega}{\zeta}\right)^{1/(2\nu)}.\label{eq:Delrms}
\end{equation}
Again, these results are true only if $\Delta_{\rm rms}\ll1$, and
again smaller $\nu$ provides for better trailing of the target.

Finally, inspired by \citet{fairhall-02}, let's examine the case of a
periodic motion of $\bar\alpha_1$ and take, for simplicity, $\bar
\alpha_1=\cA \sin \omega t$, and $v_{\bar \alpha} = \cA\omega \cos
\omega t$. Now Eq.~(\ref{eq:dD}) does not have a simple solution.
However, we search for an asymptotically periodic $\Delta(t)$ with the
same angular frequency of $\omega$.  Therefore, if we multiply
Eq.~(\ref{eq:dD}) by $\cos \omega t$, integrate over a full period,
and exchange the order of the differentiation and the integration, we
get
\begin{equation}
  \frac{d \<\Delta\,\cos \omega t \>}{dt} = -\zeta \<{\rm sign
  }(\Delta)\,|\Delta|^{2\nu-1}\cos \omega t \> - \cA\omega \<\cos^2
  \omega t\>,
\end{equation}
where $\<\dots\>$ denotes averaging over the period. Since we are
looking for stationary oscillations, time derivative applied to any
average is zero.  This gives
\begin{equation}
  \<{\rm sign }(\Delta)\, |\Delta|^{2\nu-1}\, \cos \omega t\> =-\frac{\cA\omega}{2\zeta}.
\end{equation}
Now multiplying Eq.~(\ref{eq:dD}) by ${\rm sign}(\Delta)\,
\Delta^{2\nu-1}$ and averaging again results in
\begin{equation}
  \<|\Delta|^{4\nu-2}\>= \frac{(\cA\omega)^2}{2\zeta^2},
\end{equation}
which is the same scaling as in Eq.~(\ref{eq:Dconst}).  However, now
we also have a dependence on $\omega$.

There are other cases that can be analyzed, such as a step jump in the
target, $\bar\alpha_1$, its square wave modulation, or its diffusion
in a potential (Ornstein--Uhlenbeck process). Interestingly, the last
two of these cases were used experimentally by \cite{fairhall-02}.
However, we leave the analysis for the future, when it will be
answering some specific question and won't be just a mathematical
exercise. Even with the three examples already discussed, it is clear
that letting the target move maps the scaling of the learning curve
into a stationary property (e.~g., variance of the estimation error),
which might be easier to analyze experimentally.

\section{Discussion}
\label{sec:disc}

We have shown that, with a moving target, transient learning curves
are replaced by different scaling dependences of the estimation errors
on the amplitude of the target's motion. This effect is {\em
  stationary} and may be easier to observe experimentally. However,
since we do not have a comprehensive theory of variable target
learning yet, a few precautions are in order when designing and
analyzing experiments along these lines.  (1) Target velocities must
be kept small, so that the asymptotic analysis presented in this work
holds. (2) The analysis is only valid when the learner forgets past
observations as soon as the target changes appreciably. Learning will
be much slower if such outdated samples are kept. (3) When varying the
stimulus, we have to be reasonably sure that the animal only tracks
it, but does not predict it. White noise $v_{\bar\alpha}$ or a
multiparameter representation of the target in terms of the position,
velocity, acceleration, etc., might be a solution. (4) Finally, we
have to keep in mind that, in a behaving animal, learning a change in
a signal and reacting to it may be separated by a long delay, and
special care is needed to observe the former, but not the latter.
This being said, it nevertheless is possible that all these and other
disadvantages will be outweighed by the ability to determine the
correct learning--theoretic model of the organism by varying the
amplitude and the nature (say, stochastic or periodic) of the target's
motion and studying typical responses as functions of these
parameters.

Consider, for example, the experiment described on Fig.~4 of
\citet{fairhall-02}. There the input signal (the standard deviation of
the angular velocity, $\sigma(t)$) undergoes a finite variance
$\Sigma^2$ and a finite correlation time $\tau$ random motion. The
instantaneous neuron firing rate $r(t)$ is the estimate of
$\sigma(t)$. Repeating exactly the same randomly generated stimulus
many times and averaging over spike trains, one may estimate $r(t)$
and, consequently, the rms estimation error $\Delta_{\rm
  rms}=\<(\sigma(t)-r(t)^2\>^{1/2}_t$.  Studying dependence of
$\Delta_{\rm rms}$ on $\Sigma$ and $\tau$ along the lines of
Eq.~(\ref{eq:Delrms}), one can estimate $\nu$.  Any $\nu\neq 2$
uniquely determines the underlying computational model.  For $\nu=2$,
to distinguish a usual finite parameter model from the one that is
nested, one makes the signal multidimensional (other parameters of the
angular velocity, such as the mean and the skewness, vary together
with $\sigma$). For, at least, some signal extensions, the nested
model will change the magnitude (but not the scaling) of $\Delta_{\rm
  rms}$ since $\zeta \propto 1/K(\bar r)$. In contrast, the simpler
model will keep the same prefactor but will be converging only to an
approximation of the target.

In cognitive experiments of \cite{gallistel-etal-01}, a rat was trying
to learn reward rates on different terminals and {\em match} its
foraging habits correspondingly.  It was determined to be an ideal
change detector.  Now to build a more detailed model of the animal,
one can vary the reward rates continuously, repeat experiments many
times, and then look at the average mismatch between the stimulus and
the response.  Then dependence of the mismatch on the parameters of
the rate changes will point at a proper class of learning--theoretic
models to compare the rat to.  Similarly, one can do this type of
analysis on artificial neural networks designed explicitly to model
particular animal behavior \citep{seung-03}; this will build
connections between network architectures and types of inference tasks
performed by them.

Another conclusion of our work is that the nested setup may learn
faster than the QFT one under some conditions. Thus if one desires a
complete learning machine, a nested machine should be built unless
there is some specific reason to do the opposite (such as knowing that
the world is unlikely to have sharp cutoffs).  With experiments along
the lines suggested above, this prediction should be testable.  We
should be able to see if our intuitive beliefs about appropriate
complexities of learners for particular tasks match the Nature's
choices. It would also be interesting to study if structural
characteristics of a learner are correlated with its
learning-theoretic description. That is, could it be that modular,
irregular networks, like those seen in biochemistry, often compute
like parametric or nested machines? And could layered, regular
networks in our brains, which are believed to be able to solve the
most complicated learning problems, be realizing QFT machines instead?

\appendix
\section{Model family selection in the nested setup}
\label{mfsel}

Inference in Bayesian i.\ i.\ d.\ setup is quite standard
\citep{bayes,BCS,vijay-97,bayes2,bayes3}, and nested case is not very
different. For example, a posteriori expectations of parameter values
are given by a derivative of the posterior moment generating function
(or the partition function), $Z(\bJ)$:
\begin{align}
  \< \alpha_\mu \> &= \left.\frac{\partial}{\partial
      J_\mu}\right|_{{\bf J}=0} \log \ZJ\,,\label{eq:fin_amu}\displaybreak[0]\\
  \ZJ  &\equiv \int d\balpha\, \cP(\balpha)\, {\rm
    e}^{-\cL(\balpha) + {\bf J}\cdot\balpha}\,\label{eq:ZJ}\displaybreak[0]\\
  &= \sum_r \cP(r) \ZJr{r}\,,\label{eq:ZJnested}\displaybreak[0]\\
  \ZJr{r} &\equiv \int d^{K(r)} \alpha \,
  {\rm e}^{-\cL_r(\balpha) + \sum_{\mu=1}^{K(r)}J_\mu \alpha_\mu}
  \,,\label{eq:ZJr}\displaybreak[0]\\
  \cL(\balpha) &\equiv  \sum_{i=1}^N \phi(x_i|\balpha)\,,\label{eq:L}\displaybreak[0]\\
  \cL_r(\balpha) &\equiv -\sum_{\mu=1}^{K(r)} \log
    p(\alpha_\mu) + \sum_{i=1}^N \phi(x_i|\balpha)\,,\displaybreak[0]\\
  \phi(x|\balpha) &\equiv -\log Q(x|\balpha)\,.
\end{align}
The posterior expectations are thus determined by the properties of
the $\ZJ$, which can be calculated using the saddle point analysis for
$N\gg1$.  This is difficult for the first form of $\ZJ$,
Eq.~(\ref{eq:ZJ}), due to the singularity at $\alpha_\mu =0$ [the
singularity was also the reason why we left $\cP(\balpha)$ out of the
combined Lagrangian, Eq.~(\ref{eq:L})]. Hence we return to the nested
form, Eq.~(\ref{eq:ZJnested}, \ref{eq:ZJr}), but the equivalence
between the representation should be kept in mind.  Exchanging the
order of integration and summation in Eqs.~(\ref{eq:ZJ},
\ref{eq:ZJnested}) and similar is possible if the priors decay
sufficiently fast at $r\to\infty$, or are regularized with
regularization lifted after averages are calculated. Unless mentioned
otherwise, this is always assumed.

The expectation of $\alpha_\mu$ in the model families with $K(r)<\mu$
is necessarily zero, and a similar bias towards smaller magnitudes of
parameters will be present when we average over families.  Therefore,
the a priori decrease of the variances with $\mu$,
Eqs.~(\ref{eq:finite_var}, \ref{eq:finite_varscale}), will persist a
posteriori for finite $N$. This is the famous \cite{james-stein-61}
shrinkage.

The saddle point, also called {\em classical} or {\em maximum
  likelihood}, values of parameters in each family,
$\balpha_r^{*}\equiv \{\alpha^*_{\mu;r}\}$, and the second derivatives
matrix at the saddle, $\bbF_r$, are determined by (remember that
$\alpha_{\mu;r}^{*}\equiv 0$ for $\mu>K(r)$)\footnote{The are
  possibilities of more than one saddle point and of other anomalies.
  This was analyzed by \cite{bnt-01}.  The conditions to prevent such
  problems are mild, and we assume them to hold in what follows.}
\pagebreak[0]
\begin{align}
  \left.\frac{\partial \cL_r(\balpha)}{\partial
      \alpha_\mu}\right|_{\balpha=\balpha_r^{*}} &= 0\,,\; \mu\le K(r)\,,
  \label{eq:sp}\displaybreak[0]\\
  \left.\frac{\partial^2 \cL_r(\balpha)}{\partial \alpha_\mu
      \partial\alpha_\nu} \right|_{\balpha=\balpha_r^{*}} &=
  \bbF_r^{\mu\nu}\,,\; \mu,\nu\le K(r)\,. 
\end{align}
To the first order in $1/N$, this gives
\begin{equation}
  \ZJ = \sum_r \cP(r) \frac{\cP(\balpha_r^*|r) (2\pi)^{K(r)/2}}{N^{K(r)/2}
  \det^{1/2} \frac{\bbF_r}{N}} Q(\{x\}|\balpha_r^*) \, {\rm e}^{\frac{1}{2}
  {\bf J}_r\bbF_r^{-1}{\bf J}_r + {\bf J}_r\cdot \balpha_r^*}\,,
\end{equation}
where $\mu\le K(r)$ components of ${\bf J}_r$ are the same as those of
${\bf J}$, and all higher order components are zero. Differentiating,
we get:
\begin{gather}
  \<\alpha_\mu\> = \frac{\sum_{r=1}^R \alpha_{\mu;r}^* \,{\rm e}^{-\cL(r)}} 
  {\sum_{r=1}^R {\rm e}^{-\cL(r)}}\,,\label{eq:sumr}\displaybreak[0]\\
  \cL(r) \equiv -\log \cP(r) - \sum_{\mu=1}^{K(r)}\log p(\alpha_{\mu;r}^*)
  + \sum_{i=1}^N \phi(x_i|\balpha_r^*)
    +\frac{K(r)}{2}\log\frac{N}{2\pi} + {\rm Tr}\, \log
    \frac{\bbF_r}{N}  \,.
  \label{eq:Lr}
\end{gather}
For finite $R$, and $\beta=0$, this is the usual Bayesian model family
selection: a posteriori expectations are weighted sum over posterior
probabilities of families defined by ${\rm e}^{-\cL(r)}$.  This
posterior includes the negative maximum likelihood term, $\sum_{i=1}^N
\phi(x_i|\balpha_r^*)$, which grows in magnitude linearly with $N$,
but decreased as $r$ grows due to nestedness. It also incorporates the
fluctuation determinant $\frac{K(r)}{2}\log\frac{N}{2\pi} + {\rm Tr}\,
\log \frac{\bbF_r}{N}$, which grows logarithmically in $N$, but
increases with $r$.  Depending on the value of $N$, there will be some
$r^*$, for which $\cL(r)$ is minimal.  For large $N$, as a discrete
analog of the saddle point argument, this value will dominate the sums
in Eq.~(\ref{eq:sumr}), hence some model family will be ``selected.''

However, Eqs.~(\ref{eq:sumr}, \ref{eq:Lr}) become more interesting if
one lets $R\to\infty$. The completeness condition ensures that for
large enough $r$ one will be overfitting the data, and
$Q(x|\balpha_r^*)\to 1/N \sum \delta(x-x_i)$. Therefore, if the sums
are dominated by $r\to\infty$, then consistency breaks and the
learning fails. One would thus expect two features to influence the
success of the learning. First, it is the prior $\cP(r)$, which
switches on extra degrees of freedom: for slowly decaying priors one
would expect $r\to\infty$ terms to win.  Second, it is the dependence
of the likelihood term on $r$, which measures how capable are the
newly activated degrees of freedom of overfitting, or, equivalently,
how fast $\max_i Q(x_i|\balpha_r^*)$ grows.

From Eq.~(\ref{eq:Lr}) it is easy to see that large $r$ will have an
exponentially small weight in the posterior probability if
\begin{equation}
  \lim_{r\to\infty}\frac{N \max_i \log Q(x_i|\balpha_r^*) +\log \cP(r)}
  {K(r) \log N}  =0\,.\label{eq:cond}
\end{equation}
Under this condition, $Q(x|\balpha^*)$ will eventually approach the
correct distribution, but not the sum of $\delta$--functions.
Colloquially, Eq.~(\ref{eq:cond}) requires the explanatory capacity of
the new, high order degrees of freedom to be small enough so that
keeping them always ``on'' does not make sense.  This criterion, which
we have not seen explicitly presented anywhere before, is similar to
the consistency condition of the Structural Risk Minimization (SRM)
theory, which requires that the Vapnik--Chervonenkis dimension, the
SRM capacity measure of the selected model, grows slower than the
number of samples to be explained \citep{vapnik-98,nemenman-00}.

As an example, let's analyze how the condition in Eq.~(\ref{eq:cond})
may be violated for $K(r) \sim r$.  In this case $\cP(r)$ must be
superexponential to be relevant for finding $r^*$.  Thus it is not
required to decay at some minimal speed as might have been expected,
though a need to exchange the order of integrations and summations in
arriving to Eq.~(\ref{eq:Lr}) may still force that.  Due to light
tails and small effective support, exponentially decaying priors are
not very interesting, so we disregard the prior term in
Eq.~(\ref{eq:cond}).  Then, for a fixed large $N$, a finite $r^*$ will
be dominant if $\log Q(x_i|\balpha_r^*)/r \to 0$. That is, the growth
of the $\delta$ function--like peaks of the maximum likelihood
distribution should be superlinear in $K$, the number of parameters in
the model family, in order for $r^*\to\infty$ and Bayesian setup to be
inconsistent.

\section{Fourier polynomials nested model}
\label{nmfourier}
To compare nonparametric and finite parameter nested scenarios
directly, we analyze the following example. Consider families of
probability distributions periodic on $[0,1)$, and with the logarithms
of the distributions given by Fou\-rier polynomials of degree
$r<\infty$, as in Eq.~(\ref{eq:fourier}). Due to the normalization
condition, Eq.~(\ref{eq:a0}), the number of parameters in the $r$'th
model family is $K(r)=2r$.  With an appropriate choice of priors,
Eqs.~(\ref{eq:turn_on}, \ref{eq:pr_joint}), these families form a
nested set, and the completeness for $R\to\infty$ follows from the
Fourier theorem.

The classical solution for this parameterization is ($1<\mu\le r$)
\begin{gather}
  \frac{\alpha^{*\pm}_\mu}{\sigma_\mu^2} +
  \sum_i \left(\begin{array}{c}\cos\\\sin\end{array}\right)
  2\pi\mu x_i -N \int dx\, Q(x|\balpha^*)\,
  \left(\begin{array}{c}\cos\\\pagebreak[0]\sin
    \end{array}\right) 2\pi\mu x
  \equiv\nonumber\\
  \frac{\alpha^{*\pm}_\mu}{\sigma_\mu^2}  + \frac{N}{2}\Delta^\pm_\mu
  -\frac{N}{2}Q^{*\pm}_\mu=0\,.\label{eq:spfinite}
\end{gather}
Here $Q^{\pm}_\mu$ are the cosine (sine) amplitudes of the $\mu$'th
mode in the Fourier expansion of $Q(x|\balpha)$, and
$\Delta^{\pm}_\mu$ are the same for the empirical probability density,
$1/N\sum \delta(x-x_i)$. $\Delta^{\pm}_\mu$ are also the stochastic
Fourier transform of $Q(x)$. The cosine--cosine components of the
second derivative matrix at the saddle point are 
\begin{eqnarray}
  \left.\frac{\partial^2 \cL}{\partial \alpha^{+}_\mu \partial
    \alpha^{+}_\nu}\right|_{\balpha=\balpha^*}
  &=& \frac{\delta_{\mu\nu}}{\sigma^2_\mu} + N\int dx\, Q(x|\balpha^*)
  \, \cos
  2\pi\mu x \, \cos
  2\pi\nu x \nonumber\\
  &&- N \int dx\, Q(x|\balpha^*)\,
  \cos 2\pi\mu x \, \int dy\, Q(y|\balpha^*)\,
  \cos 2\pi\nu y  \\
  &=& \frac{\delta_{\mu\nu}}{\sigma^2_\mu} + \frac{N}{4}(Q^{*+}_{\mu+\nu} +
  Q^{*+}_{\mu-\nu})+
  \frac{N}{2}Q^{*+}_\mu 
  Q^{*+}_\nu\;,
\end{eqnarray}
and the sine--sine and the sine--cosine components are written
similarly.  This matrix is provably positive definite. Thus for
$N\to\infty$ we can perform the saddle point analysis.

For $\beta=0$, the variance $\sigma^2_\mu$ is constant, and we can
neglect the first term in Eq.~(\ref{eq:spfinite}) in the limit of
large $N$.  This leads to the following solution of the saddle point
equations:
\begin{equation}
  Q^{*\pm}_\mu \approx \Delta^{\pm}_\mu,\quad \quad \mu=1\dots r.
  \label{eq:sol}
\end{equation}
For $\beta>0$, $Q^{*\pm}_\mu$ will be corrected by a systematic
$\beta$--dependent bias, which will tend to 0 for fixed $\mu$ as $N$
grows. This will decrease the posterior variance of the estimator
$Q^*$.

Equation (\ref{eq:sol}) says that the first $r$ pairs of coefficients
$\alpha^{*\pm}_\mu$ are such that the corresponding Fourier amplitudes
of the classical solution $Q^*$ match those of the empirical one. By
Nyquist theorem and the law of large numbers, for $r<N/2$,
$\Delta^\pm_\mu$ approach the Fourier amplitudes of the unknown target
probability density $\bar Q $. Thus the low frequency modes will be
learned well.  However, if $r >N/2$ the saddle point solution will
start to overfit and develop $\delta$--like spikes at each observed
data point. This is in accord with the observation we have already
mentioned: to guarantee consistency, the capacity of models, as
measured by either the VC dimension or the scaling dimension of
\cite{bnt-01}, which in this case is equal to the number of free
parameters, must grow slower than $N$.

To avoid overfitting when averaging over $r$, we must make sure that
the contribution of $r\to\infty$ to the posterior log--probability,
Eq.~(\ref{eq:Lr}), is negligible. In this regime, according to
Eq.~(\ref{eq:sol}), the $r$ available modes will create peaks of
height $\sim r$ (recall the Fourier expansion of the
$\delta$--function) at the observed sample points. With $K(r)=2r$,
this ensures consistency by satisfying Eq.~(\ref{eq:cond}).

Further, we can prove that $r^*$ is not only finite, but actually
grows sublinearly in $N$, again paralleling results for SRM
\citep{vapnik-98} and their Bayesian equivalent \citep{nemenman-00}.
Suppose $r\gg N$ dominates the posterior. Then, for a slowly decaying
$\cP(r)$, Eq.~(\ref{eq:Lr}) can be rewritten as
\begin{equation}
  \cL(r) \sim -N\log r + r\log N \,,
\end{equation}
This is minimized (and the posterior probability is maximized) for
$r^*\sim\format{N/\log N}{\frac{N}{\log N}}$, and higher values of $r$
are exponentially inhibited.  Thus the assumption of $r\gg N$ being
dominant is incorrect, and the posterior probability is dominated by
$r^*\lesssim N$ for all reasonable priors.  This is, of course, the
worst case estimation, and in many typical applications the value of
$r^*$ is even lower.


\section{Fourier nested model and QFT targets}
\label{crossNestQFT}
As shown above, $r^*$ that minimizes Eq.~(\ref{eq:Lr}) for a Fourier
nested setup is much smaller than $N$. This is true for any target,
including QFT--typical targets.  For $r$ of such magnitude, the first
$r$ modes of the target are well approximated by the estimate, and
they contribute $O(r/N)$ to the leading data dependent term in
Eq.~(\ref{eq:Lr}). The modes of the target above the $r$'th are not
fitted by the estimate, and each of them contributes its variance of
about $\ell^{-2\eta+1}\mu^{-2\eta}$ to the data dependent term, adding
up to $\sum_{\mu=r+1}^{\infty} \ell^{-2\eta+1}\mu^{-2\eta} \propto
(r\ell)^{-2\eta +1}$.  Combined with the fluctuation determinant this
gives
\begin{equation}
  \cL(r) \sim - N (r\ell)^{-2\eta+1} + r \log N
\end{equation}
for determining the most probable $r$. Thus for $N\gg1$ (or
$\Lambda\ll1$)
\begin{eqnarray}
  r^* &\propto& \left(\frac{N}{\log N}\right)^{1/2\eta}\ell^{1/2\eta-1}\,, \mbox{ and}\\
  \cD&\propto& N^{1/2\eta} \left(\frac{\log N}{\ell}\right)^{1-1/2\eta}\,.\label{eq:Dcross1}\\
  \Lambda &\sim& \left(\frac{\log N}{N\ell}\right)^{1-1/2\eta}\,.\label{eq:Lcross}
\end{eqnarray}
Due to the many simplifications made here, the exact form of the
logarithmic terms in these expression is questionable,\footnote{If
  certain derivative of the target distribution satisfies some
  Lipschitz conditions, then the Occam factor and the learning curve
  for histogramming density estimators provably have logarithmic
  contributions \citep{hall-hannan-88,rissanen-etal-92}.  In contrast,
  logarithmic corrections for QFT models and for parametric learning
  of QFT--typical targets have not yet been analyzed.  However, the
  logarithmic differences between the cases have been expected: in
  discrete case, $\beta=0$, once we know $K^*\sim N^\omega$, each of
  $K^*$ parameters is free to vary with the same variance, giving
  familiar $N^\omega\log N$ fluctuations.  For the nonparametric case,
  $\sigma_\mu<\sigma_\nu$ for $\mu>\nu$.  Thus each next parameter
  varies less, somewhat decreasing the total fluctuations
  \citep{bnt-01}.
  
  These logarithmic terms have the same roots as the difference
  between cross--validation, bootstrap, and Akaike's model selection
  criterion on one hand and Dawid's prequential statistics and
  Bayesian model selection on the other \citep{stone-77,dawid-84}.
  There the difference in the magnitude of the prediction error is
  also due to most of the parameters that are active at a given $N$
  being latent for smaller sample sizes.} and, in practice, they are
impossible to observe for realistic $N$ due to the target--dependent
prefactors in front of the universal scaling term and various
statistical fluctuations.  However, the power law in
Eq.~(\ref{eq:Lcross}), which is definitely correct, suggests that the
performance of the nested model is comparable to that of the true QFT
one. In particular, the nested learning machine also can solve
arbitrarily complex inference problems.

A rigorous way to estimate performance of the nested learning on a
nonparametric target is to calculate $\<\rho(\epsilon)\>= \int d
\bbalpha \, \cP(\bbalpha) \rho(\epsilon; \bbalpha)$, where $\rho$ is
of the form Eq.~(\ref{eq:rhofinsum}), and the averaging is done over
the QFT prior, and then calculate $\cD$ and $\Lambda$ from this
averaged $\rho$. This is difficult, and instead we may choose to
replace $\<\rho(\epsilon)\>$ by $\rho(\epsilon;\bbalpha_{\rm typ})$,
where $\bbalpha_{\rm typ}$ is a typical target in the nonparametric
prior, Eq.~(\ref{eq:npprior}).\footnote{The benefit
  $\<\rho(\epsilon)\>$ provides over $\rho(\epsilon;\bbalpha_{\rm
    typ})$ is knowing the prefactors in $\cD$ and $\Lambda$.  We don't
  believe that any of the priors studied in this work will be {\em
    exactly} realized in nature. Therefore, calculation of
  $\<\rho(\epsilon)\>$ is not a priority.}  For such $\bbalpha_{\rm
  typ}$, $D_r(\bbalpha) \sim \sum_{\mu>r} \ell^{-2\eta+1}\mu^{-2\eta}
\propto (\ell r)^{-2\eta+1}$.  Further, $\cP(\hat\alpha^{\pm}_{\mu;
  {\rm typ}}|r) \sim \exp [-0.5\,
\mu^{-2\eta}\ell^{-2\eta+1}/\sigma^2_\mu]$. In our case, $\sigma^2_\mu
\sim \mu^{-2\beta}$. Therefore, for $2(\eta-\beta)>1$, which is
satisfied for $\eta>1/2$ and $\beta=0$, this gives
$\cP(\hat\balpha_r|r) \sim \exp \left[-\sum_{\mu=1}^r
  \mu^{-2\eta}\ell^{-2\eta+1}/\sigma^2_\mu\right] \sim \exp\left[ C_1
  - C_2 r^{-2(\eta-\beta)+1}\right]$, where $C_1$ and $C_2$ are
constants.  For large enough $r$, this whole expression tends to a
constant. Thus, combining with Eq.~(\ref{eq:rhofinsum}), we get
\pagebreak[0]
\begin{eqnarray}
    \rho(\epsilon;\bar\balpha_{\rm typ})
  &\sim&\sum_{r:\,r^{-2\eta+1}\le\epsilon} \cP(r)
 \frac{2 \pi^{r}}{\Gamma(r)} 
  \frac{ [\epsilon-(\ell r)^{-2\eta+1}]^{(r-1)}}{\sqrt{\det \cF_{K(r)}}}\,.
  \label{eq:rhocross1}
\end{eqnarray}
If $\cP(r)$ is subexponential as before, we get $\rho$ to the leading
order in small $\epsilon$ by calculating the sum in
Eq.~(\ref{eq:rhocross1}) using the saddle point analysis and taking
just the zeroth order term. The saddle value for $r$ is $r^*\sim
\epsilon^{-1/(2\eta -1)}\ell^{-1}$, which gives
\begin{equation}
    \rho(\epsilon;\bar\balpha_{\rm typical}) 
    \sim \epsilon^{\epsilon^{-1/(2\eta-1)}\ell^{-1}}\,,
\end{equation}
with the first subleading term of
$O\left(\exp\left[-\epsilon^{-1/(2\eta-1)}\ell^{-1}\right]\right)$.
Doing the leading order evaluation of the integral in
Eq.~(\ref{eq:D}), we now get $\epsilon^* \sim \left(N\ell/\log
  N\right)^{1/2\eta -1 }$, which again results in
Eq.~(\ref{eq:Dcross1}).

In summary, learning a distribution typical in the nonparametric
model by means of the nested setup results in, at most, a logarithmic
performance loss. 

\section*{Acknowledgments}
I thank William Bialek and Chris Wiggins for many stimulating
discussions. I am also grateful to Ila Fiete and two anonymous
referees for carefully reading this manuscript and providing important
feedback. This work was supported in part by NSF grants PHY99--07949
to Kavli Institute for Theoretical Physics and ECS-0332479 to Chris
Wiggins and myself.

\bibliographystyle{plainnat} \bibliography{ic_param}
\end{document}